\documentclass{aa}

\newcommand{\xte}{XTE J1118+480 }
\newcommand{\xtee}{XTE J1118+480}

\usepackage{epsfig}
\usepackage{graphics}
\usepackage{subfigure}
\usepackage{natbib}
\usepackage{amsmath}
\usepackage{enumerate}

\begin{document}

\title{A model for the polarization of the high-energy radiation from accreting black holes: the case of XTE J1118+480}

\author{F. L. Vieyro\inst{1,2,3}, G. E. Romero\inst{1,2} \and S. Chaty\inst{3,4}}
  
\institute{Instituto Argentino de Radioastronom\'{\i}a (IAR, CCT La Plata, CONICET), C.C.5, (1984) Villa Elisa, Buenos Aires, Argentina \and Facultad de Ciencias Astron\'omicas y Geof\'{\i}sicas, Universidad Nacional de La Plata, Paseo del Bosque s/n, 1900, La Plata, Argentina \and AIM, UMR-E 9005 CEA/DSM-CNRS-Universit\'e Paris Diderot, Irfu/Service d'Astrophysique, Centre de Saclay, 91191 Gif-sur-Yvette Cedex, France \and Institut Universitaire de France, 103 boulevard Saint-Michel, 75005 Paris, France}

\offprints{F. L. Vieyro \\ \email{fvieyro@iar-conicet.gov.ar}}

\titlerunning{Polarization in accreting black holes}

\authorrunning{Vieyro, et al.}

\abstract
{The high-energy emission ($400$ keV $- 2$ MeV) of Cygnus X-1 --the most well-studied Galactic black hole-- was recently found to be strongly polarized. The origin of this radiation is still unknown.} 
{In this work, we suggest that it is the result of non-thermal processes in the hot corona around the accreting compact object, and study the polarization of high-energy radiation expected for black hole binaries. }
{Two contributions to the total magnetic field are taken into account in our study, a small scale random component related to the corona, and an ordered magnetic field associated with the accretion disk. The degree of polarization of gamma-ray emission for this particular geometry is estimated, as well as the angle of the polarization vector. }
{We obtain that the corona+disk configuration studied in this work can account for the high degree of polarization of gamma-rays detected in galactic black holes without the need of a relativistic jet; specific predictions are made for sources in a low-hard state. In particular, the model is applied to the transient source \xtee; we show that if a new outburst of \xte is observed, then its gamma-ray polarization should be measurable by future instruments, such as ASTRO-H or the proposed ASTROGAM. }
{}

 
\maketitle

\section{Introduction}

The hard X-ray/soft gamma-ray spectra of several black holes binaries (BHBs) present two components: a power-law component with an exponential cutoff, which is the result of Comptonization of soft photons from the disk by a hot corona, and a steeper non-thermal spectrum up to MeV energies \citep[e.g.,][]{McConnell2000,ling2003,ling2005,cadolle2006}. The detection of a high degree of polarization ($67\pm30$ \%), in the non-thermal tail of the high-mass X-ray binary Cygnus X-1, confirmed that these components have different emission processes \citep{laurent2011,jourdain2011,romeroVieyro2014}.
 
Different scenarios have been proposed to explain the origin of the high energy component \citep{poutanenCoppi1998,zdziarski2012,zdziarski2013,flor01,vieyro2012}. Ultimately, its nature still remains unknown.

The hybrid thermal/non-thermal corona model of \citet{poutanenCoppi1998} attributes the high-energy tail of BHBs to Comptonization of the soft photons of the disk by a non-thermal population of electrons. This model has been successful in reproducing the soft gamma-ray spectrum of accreting black holes; however, Compton scattering is not consistent with the high polarization fraction measured by  {\it INTEGRAL} \citep{laurent2011}.

Highly polarized radiation must be produced in the presence of an ordered magnetic field. A relativistic jet is, then, the first candidate for the origin of the gamma-ray emission. \citet{zdziarski2012}  proposed that the soft gamma-ray radiation is generated in a relativistic jet (see also \citealt{zdziarski2014,zdziarski2013}). They studied synchrotron and self-Compton emission due to non-thermal electrons. In order to reproduce the MeV component, they require an electron spectrum that is difficult to explain with diffusive acceleration processes. For models with more realistic particle distributions, the synchrotron fluxes at $E\sim$ MeV  are below those observed in Cygnus X-1. In a recent work by \citet{pepe2015}, good fits to the MeV emission are obtained for several sets of parameters. Nevertheless, current models for the polarization of the jet present problems to reproduce the polarization angle measured in Cygnus X-1. 


The studies by \citet{flor01} and \cite{vieyro2012} of the transport of non-thermal particles in the corona of Cygnus X-1 have successfully reproduced the emission detected by {\it COMPTEL} and  {\it INTEGRAL} at MeV energies (see also \citealt{vieyroGRO2012}).  Furthermore, \citet{romeroVieyro2014} showed that the same radiation from the corona can develop the high degree of polarization detected by \citet{laurent2011}. In this context, the synchrotron emission of secondary leptons is responsible for the polarized radiation.

In what follows, we analyze the influence of a given configuration of the magnetic field in the polarization of synchrotron radiation, expanding our previous work on the topic \citep{romeroVieyro2014}. We then apply the model to \xtee. This is a transient source that was discovered during an outburst in 2000 \citep{remillard2000}, and it was later detected in a second outburst in 2005 \citep{zurita2005}. Its location at high latitude over the Galactic plane --where the interstellar absorption is low-- allowed several multi-wavelength studies of both events \citep{chaty2003,maitra2009,brocksopp2010}. As a result, its spectrum was well characterized, and in both occasions it resembled the low-hard state of BHBs.

Our article has the following structure: in Sect. \ref{model} the basic hypothesis of the model are stated. We study the non-thermal particle interactions in the corona, and compute the final spectral energy distribution (SED). In Sect. \ref{pol}, we describe a specific geometry of the magnetic field and estimate the polarization degree and the polarization vector angle. In Sect. \ref{xte} the case of \xte is considered. We present our results in Sect. \ref{results}, discuss them in Sect. \ref{discussion}, and finally conclude in Sect. \ref{conclusions}.

\section{Basic model}\label{model}

We consider a spherical corona with a radius $R_{\rm{c}}$, and an accretion disk that is truncated at $r_{\rm{in}}$. The ratio $r_{\rm{in}}/R_{\rm{c}}$ is taken as $0.9$, which is a standard value for the low-hard state \citep{poutanen1998}. The corona is assumed to be magnetically dominated and static, where particles can be removed by diffusion \citep{bisnovatyi1977}.

The hard X-ray photon density energy distribution of the corona, $n_{\rm{ph}}$, is represented as a power-law with an exponential cutoff $E_{\rm{c}}$,

\begin{equation}
n_{\rm{ph}} (E_{\rm{ph}}) \propto E_{\rm{ph}}^{\alpha} e^{-E_{\rm{ph}}/E_{\rm{c}}}.
\end{equation}

\noindent The normalization constant can be obtained from the bolometric luminosity according to:

\begin{equation}
\frac{L_{\rm{c}}}{4\pi R_{\rm{c}}^2 c } = \int_0^{\infty} E_{\rm{ph}} n_{\rm{ph}} (E_{\rm{ph}}) dE_{\rm{ph}}.
\end{equation}

The accretion disk is modeled as a multi-temperature black body, as described in \citet{frank2002} and \citet{vila2012}. At every radius, the disk is in local thermal equilibrium, and it radiates as a black body of temperature $T(r)$; the temperature profile for a standard disk is given by \citet{shakura1973}:

\begin{equation}
T(r) = \frac{T_{\rm{max}}}{0.488} \Big( \frac{r}{r_{\rm{in}}} \Big)^{-3/4} \Big( 1- \sqrt{\frac{r_{\rm{in}}}{r}} \Big)^{1/4} ,
\end{equation}

\noindent where $T_{\rm{max}}$ is the maximum temperature of the disk. The integrated disk spectrum can be obtained as

\begin{equation}
F_{\gamma} (E_{\gamma}) = 2\pi \frac{\cos i}{d^2} \int^{r_{\rm{out}}}_{r_{\rm{in}}} B(E_{\gamma},T(r)) r dr,
\end{equation}

\noindent where $i$ is the inclination angle (we consider that the accretion disk is parallel to the orbital plane), and $B(E_{\gamma},T(r))$ is the Planck function, given by

\begin{equation}
B(E_{\gamma},T(r)) = \frac{2}{c^2 h^3} \frac{E_{\gamma}^3}{\exp[E_{\gamma}/kT(r)] -1}.
\end{equation}

The low-hard state of BHBs is usually associated with the presence of relativistic jets. These jets are thought to be launched by a magnetic mechanism \citep[e.g.,][]{blandford1982}. Since in several systems the jet kinetic power is of approximately the same luminosity as the corona, we estimate the value of the random magnetic field in the corona by assuming equipartition between the magnetic energy density and the bolometric photon density of the corona \citep{bednarek2007}, 

\begin{equation}
		\frac{B_{\rm{c}}^2}{8\pi}=\frac{L_{\rm{c}}}{4\pi R_{\rm{c}}^2c}.
	\end{equation}
	
For a more detailed description of the corona model the reader is referred to \citet{flor01} and \citet{vieyro2012}.

\subsection{Non-thermal particle injection}

A likely mechanism of particle acceleration in accreting black hole coronae  is magnetic reconnection \citep[e.g.,][]{bette2010,lazarian2011}. In this process particles undergo a Fermi-type acceleration in magnetic reconnection sites, which are regions where the topology of the magnetic field change as a result of magnetic fluxes of opposed polarity approaching each other. We consider an efficiency for this mechanism of $\eta \sim 0.01$ \citep{vieyro2012}. The resulting particle spectrum for a diffusive acceleration mechanism is a power-law. Accordingly, we adopt an injection function of primary particles given by:

\begin{equation}
Q(E) \propto E^{-\Gamma} e^{-E/E_{\max}},
\end{equation}

\noindent where $E_{\max}$ is the maximum energy, and it is determined by a balance between the cooling rates and the acceleration rate. The normalization constant depends on the available energy to accelerate particles, $L_{\rm{rel}}$. We adopt $q = L_{\rm{rel}}/L_{\rm{c}}=0.1$, a value within a reasonable range in the parameter space \citep{delValle2011,vieyro2012}. The way in which the energy is distributed between hadrons and leptons is unknown. It is useful to define the parameter $a$ as the ratio of power injected in protons and electrons, that is $a=L_p/L_e$. We adopt $a=100$, as observed in galactic cosmic rays \citep{ginzburg}. The value of $a$, however, does not affect the final particle distribution index, having no effect on the estimates of the polarization degree.

\subsection{Particle transport}

Once particles are injected, they interact with the different local fields in the source (i.e., photon, matter, and magnetic fields). We study the transport of primary and secondary (pions, muons, and electron/positron pairs) particles, as well as the transport of photons, using the numerical method described in \citet{vieyro2012}.



\begin{table}
    \caption[]{Main parameters of the corona and the disk.}
   	\label{table0}
   	\centering
\begin{tabular}{lc}
\hline\hline
Parameters & Value\\ [0.01cm]
\hline 
$L_{\rm{c}}$: corona luminosity [erg s$^{-1}$]                          & $10^{36}$	 	\\
$kT_{\rm{max}}$:  maximum disk temperature [eV]								  & $22.0$   \\
$r_{\rm{in}}$:   inner disk radius [$r_{\rm{g}}$] 							& $160.0$ 	         	\\
$r_{\rm{in}}/R_{\rm{c}}$: inner disk/corona ratio               & $0.9$								\\
$\alpha$: 			X-ray spectrum power-law index    							& $1.75$								\\
$E_{\rm{c}}$:   X-ray spectrum cut-off [keV]							& $200$   			\\
$B_{\rm{c}}$: 	magnetic field [G]				 											& $10^4$		\\
$n_{i},n{e}$:   plasma density [cm$^{-3}$] 											& $10^{11}$	\\
\hline  \\[-0.5cm]
\end{tabular}
\end{table}

\begin{figure}[ht]
\begin{center}
\includegraphics[width=0.8\linewidth]{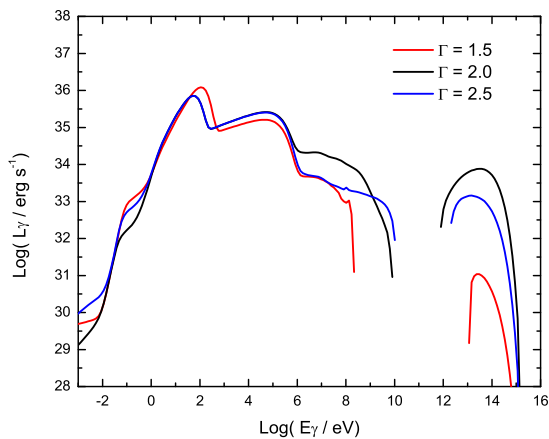}
\caption{Spectral energy distributions for different values of the primary injection index.}
\label{fig:SEDindex}
\end{center}
\end{figure}

Figure  \ref{fig:SEDindex} shows the final SED, computed for three values of the injection index $\Gamma$: the standard value $\Gamma = 2$, and $\Gamma = 1.5, 2.5$ which are intermediate values that can be obtained with a mechanism based on magnetic reconnection acceleration \citep{drury2012,bosch-ramon2012}.

The gap that appears at $\sim 10^8$ eV $< E_{\gamma} < 10^{13} $ eV is own to absorption in the internal field of the corona. Absorption by photon annihilation has a fundamental role in our model; on the one hand it shapes the final SED at MeV energies, and, on the other hand, it is the main source of secondary pairs in the corona.

The polarized radiation measured by {\it INTEGRAL} was at $400$ keV $< E_{\gamma} < 2$ MeV; the main contribution to the final flux in this energy range is the synchrotron emission of secondary pairs.

\section{Polarization}\label{pol}


An isotropic distribution of relativistic leptons with a power-law energy spectrum $ N(E) \propto E^{-\Gamma} $ in a region with a large-scale magnetic field, produces a synchrotron emission that is linearly polarized \citep{korchakov1962}. In the presence of a uniform magnetic field $B_{\rm u}$, the degree of linear polarization is \citep[e.g.,][]{pacholczyk1970}

\begin{equation}\label{eq:pol}
P_0(\Gamma) = \frac{\Gamma+ 1}{\Gamma + 7/3}.
\end{equation}

\begin{figure}[ht]
\begin{center}
\includegraphics[width=0.8\linewidth]{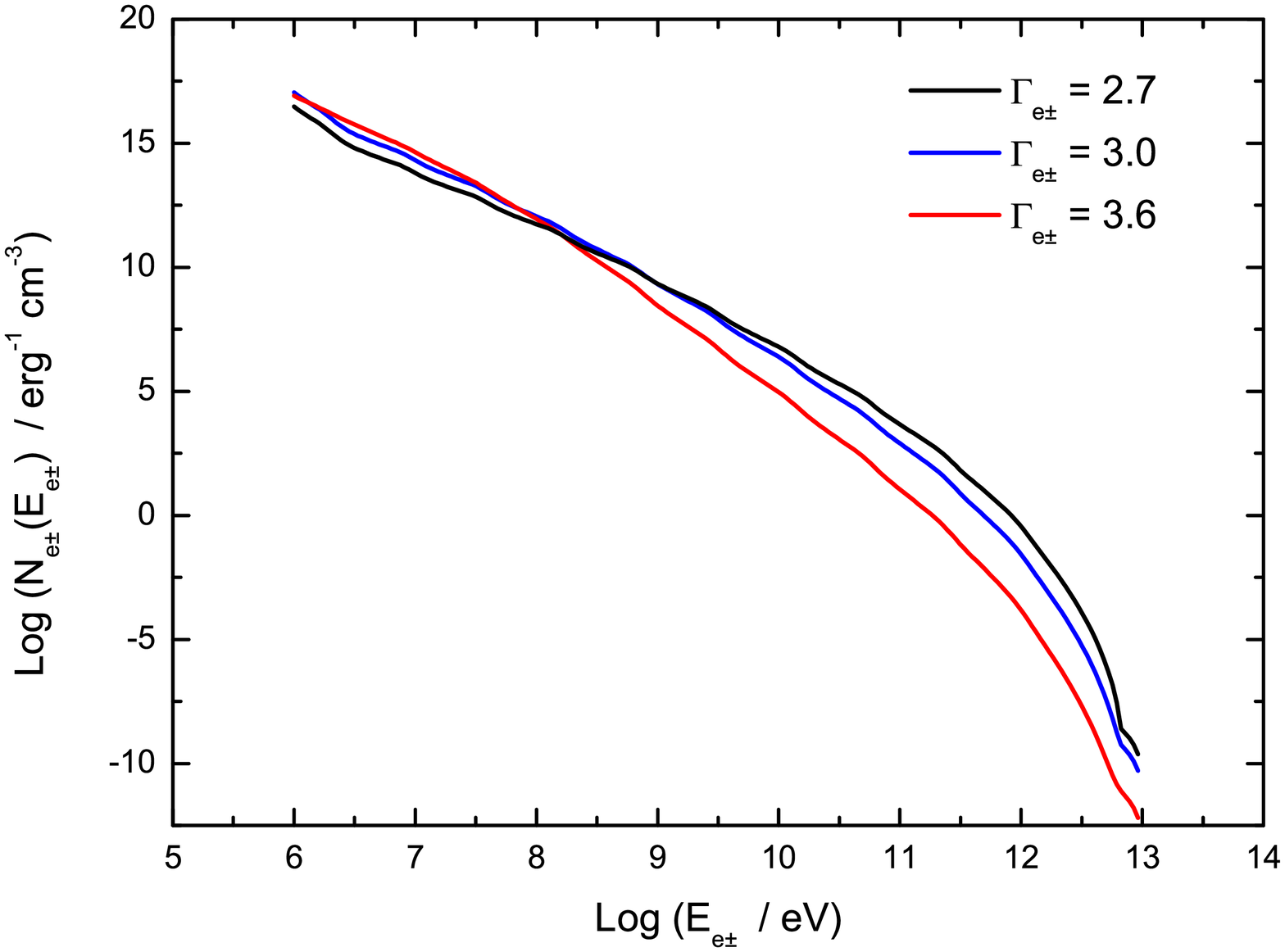}
\caption{Secondary pair distributions for the different values of the primary injection index.}
\label{fig:pairIndex}
\end{center}
\end{figure}

\noindent As an example, Fig. \ref{fig:pairIndex} shows the secondary pair distributions corresponding to the spectra of Fig. \ref{fig:SEDindex}. For a primary injection index of $\Gamma = 1.5,2,2.5$, the secondary pair index is $\Gamma_{e^{\pm}} = 2.7,3.0,$ and $3.6$, respectively. If the magnetic field were completely uniform, the polarization degree should be $73-77$\%.

However, this is an upper limit to the degree of polarization, since the non-uniformity of the magnetic field tends to decrease the value. For a magnetic field with a random component $B_{\rm{c}}$, and a spatial scale of variation smaller than the source size, the degree of polarization is reduced to \citep{burn1966}:

\begin{equation}
P (\Gamma) = P_0(\Gamma)\frac{B_{\rm u}^2}{B_{\rm u}^2+B_{\rm{c}}^2}.
\end{equation}

In BHBs, the random magnetic field is associated with the corona, whereas the uniform component is related to the accretion disk. Since the disk only penetrates the corona up to a certain radius, we consider that $B_{\rm u} = \xi B_{\rm{d}}$, where $B_{\rm{d}}$ is the magnetic field of the disk, and $\xi$ is a filling factor that depends on the fraction of the corona volume covered by $B_{\rm{d}}$. In zeroth-order approximation, $\xi = 1-  (r_{\rm{in}}/R_{\rm{c}})^3$.

\begin{figure}[ht]
\begin{center}
\includegraphics[width=0.8\linewidth]{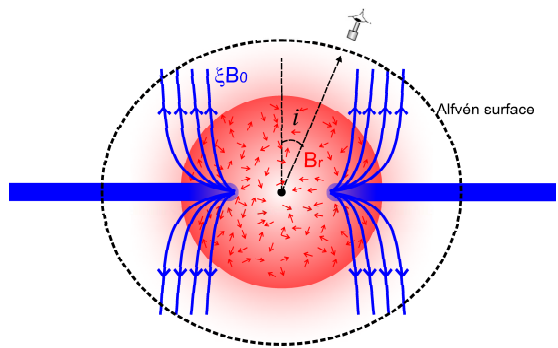}
\caption{Illustration of the geometry of the magnetic field components.}
\label{fig:Bcorona}
\end{center}
\end{figure}

We also consider that the angular distribution of the magnetic field, $B_{\rm u}(\Omega)$, has axial symmetry  (see Fig. \ref{fig:Bcorona} for a schematic representation). If $B_{\parallel}$ and $B_{\perp}$ are the components of the magnetic field of the disk parallel and normal to the symmetry axis, the observed degree of polarization can be obtained as \citep{korchakov1962}

\begin{equation}\label{eq:ptot}
P_{\rm tot}(\Gamma) = P_0 \frac{15}{8} \frac{\Gamma+5}{\Gamma+7}  \frac{B_{\rm u}^2}{B_{\rm u}^2+ B_{\rm c}^2} \frac{\left|\left\langle \Delta B^2 \right\rangle\right|}{B_{\rm u}^2} \sin ^2 i,
\end{equation}

\noindent where $i$ is the angle between the symmetry axis and the line of sight\footnote{In a previous work by \citet{romeroVieyro2014}, the factor $\sin ^2 i$ was not included in Eq. \ref{eq:ptot}. This equation was used to constrain the ratio $B_{\rm u}/B_{\rm r}$ in the low hard state. Although this ratio is affected by the missing factor, the main result of expecting higher polarization degree in an intermediate state than in the low-hard state remains valid.},  $\left\langle \Delta B^2\right\rangle= \left\langle B_{\perp}^2 \right\rangle-\left\langle B_{\parallel}^2 \right\rangle$, and the averaging is carried over out as

\begin{equation}\label{eq:promedio}
\left\langle B \right\rangle = \frac{1}{4\pi}\int{B_{\rm u} (\Omega) d\Omega}.
\end{equation}

For the magnetic field of the accretion disk, we use the geometry described in \citet{stepanovs2014}. The potential vector is defined as $\vec{A}=A\breve{e}_{\phi}$, where

\begin{equation}\label{eq:potVector}
A=\frac{4}{3}\frac{B_0}{r^{1/4}} \frac{1}{[1+4\cot^2 \theta]^{5/8}}.
\end{equation}


\noindent The magnetic field is $\vec{B}=\vec{\nabla} \times \vec{A}$, and its components are

\begin{equation}\label{eq:magneticField}
\begin{aligned}
B_{r} &=  \frac{4 B_0 }{3r^{5/4} } \frac{\cot \theta}{  \Big(1+4 \cot^2 \theta \Big)^{5/8}} \left[1+\frac{5\Big(1+\cot^2 \theta \Big)}{\Big(1+4 \cot^2 \theta \Big)} \right] ,  \\
B_{\theta} &=  -\frac{B_0}{r^{5/4}} \Big(1+4 \cot^2 \theta \Big)^{-5/8} ,  \\
B_{\phi} & =0, 
\end{aligned}
\end{equation}


\noindent where $B_0$ is obtained through the condition $B(r_{\rm{in}}, \pi/2) = B_{\rm{d}}$. The geometry represented by Eq. \ref{eq:magneticField} fulfills the necessary conditions for the magnetic field to launch an outflow. Inside the Alfv\'en region, the force-free approximation holds, and the field lines corotate with the disk. Only outside the  Alfv\'en radius the azimuthal component is relevant, since the force-free approximation is no longer valid. We adopt, then, an azimuthal component of the magnetic field equal to zero in the inner region \citep[e.g.,][]{spruit2010}.

\begin{figure}[ht]
\centering
{\includegraphics[width=0.45 \textwidth,keepaspectratio]{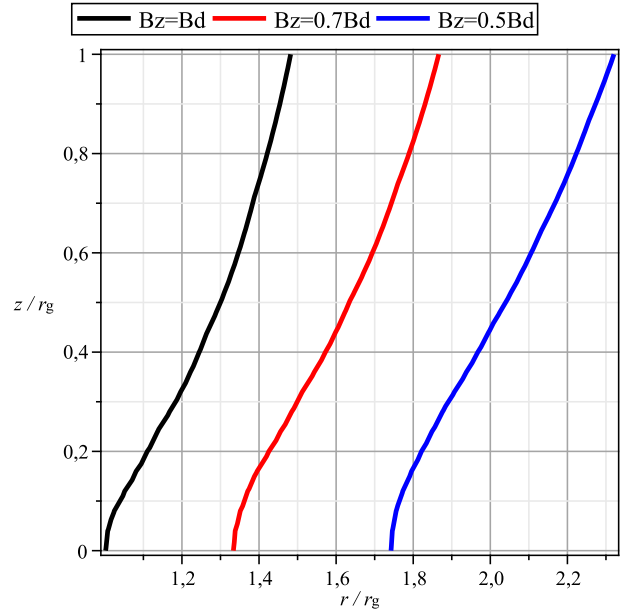}} \hspace{20pt} 
\caption{Parallel component of the magnetic field associated with the accretion disk. Both axis are normalized to $r_{\rm in}$, and $B_{\rm d}=B_{\rm d}(r=r_{\rm in},\theta= \pi/2)$.}
\label{fig:Blines}
\end{figure}

To estimate the parallel and perpendicular components of the magnetic field, we first transform the spherical components given by Eq. \ref{eq:magneticField} to Cartesian components, and then to cylindrical ones. Figure \ref{fig:Blines} shows the contour lines of the resulting parallel component of the field.


The launching of an outflow requires a minimum inclination angle of the field lines with respect to the symmetry axis of $\theta = 30^{\circ}$ \citep{blandford1982}. Therefore the quantity $\left\langle \Delta B^2 \right\rangle$ is estimated using Eq. \ref{eq:promedio}, with $30^{\circ} < \theta < 90^{\circ}$, and results in $\left\langle \Delta B^2 \right\rangle / B_{\rm u}^2 \sim 0.23$. 

The factor $\left\langle \Delta B^2 \right\rangle / B_{\rm u}^2$ decreases when the inclination of the $B_{z}$ component increases. Then, for a magnetic field that is dragged by the accretion disk to the compact object, the inclination of the field lines becomes larger, resulting in an increment of the polarization degree. This effect could be important in the high-soft state.

On the other hand, the polarization degree has a strong dependence with the inclination angle of the binary system: for an edge-on system (i.e., $i = 0$) we expect the maximum possible degree of polarization, whereas for a face-on disk the degree of polarization should be negligible.

\section{Application to \xte}\label{xte}

\subsection{Multi-wavelength observations of \xte in outbursts}\label{outburst}

The soft X-ray transient XTE\,J1118+480 was discovered by {\it RXTE} on 2000 March 29 at the Galactic coordinates ($l,b$) = ($157.62\deg$,$+62.32\deg$) as a weak (39 mCrab), slowly rising X-ray source \citep{remillard2000}; the post-analysis revealed an outburst in January 2000, with a similar brightness.

XTE\,J1118+480 location at an unusually high galactic latitude in the direction to the Lockman Hole implied a very low absorption along the line of sight ($N_{H} \sim 0.80-1.30 \times 10^{20}$ cm$^{-2}$, \citealt{chaty2003}). This low interstellar absorption allowed to get an unprecedented wealth of multi-wavelength coverage from radio to X-ray domains, including even the first extreme ultraviolet spectrum of an X-ray transient (see \citealt{chaty2003} and references therein).

Optical observations of the source in quiescence led to the determination of a large value of the mass function, f(M) $= 5.9 \pm 0.4$ M$_{\odot}$, making this source a strong black hole candidate \citep{wagner2001, mcclintock2001}. The companion star is a low mass star of $M = 0.27 \pm 0.05 M_{\odot}$, and a spectral type between K5V and K8V \citep{chaty2003,gallo2014}. XTE\,J1118+480 exhibits an orbital period of 4.1\,hr = 0.17082\,d, one of the shortest  among all black hole candidates \citep{cook2000}.

The distance to the source is estimated as $d = 1.72$ kpc, and the inclination angle of the disk with respect to the line of sight is $\sim 70^{\circ}$ \citep{gelino2006}.

This black hole X-ray transient displayed two outbursts: the first one in 2000, and the second one in 2005. Thanks to the low interstellar absorption, it was observed each time from radio to hard X-rays. In the low-hard X-ray spectral state the source exhibited correlated X-ray and radio behavior during both outbursts. The light curves, however, showed different behaviors namely. 

i.) During the 2000 outburst the source stayed in a very low low-hard state, with an inner radius of the accretion disk estimated at $\sim 350 R_s$ ($R_s$ = Schwarzschild radius for an object of mass $M$), and a strong non-thermal (synchrotron) contribution in optical and near-infrared \citep{chaty2003}. The SED of the source (from radio to X-rays) remained barely unchanged for almost three months. This long plateau-like phase was probably due to a stable jet.

ii.) The 2005 outburst was more typical of a canonical soft X-ray transient, with a more substantial contribution from the accretion disk, and a {\it fast rise, exponential decay} (FRED), light-curve (see \citealt{zurita2005,brocksopp2010} and references therein).

The different characteristics of these two outbursts make XTE\,J1118+480 a very promising source to study polarization and to disentangle its origin.

\subsection{Corona of \xte}\label{corona}

We characterize the hard X-ray emission of the corona during the 2000 outburst, as a power-law of index $\alpha \sim 1.75$ and an exponential cutoff at $E_{\rm{c}} = 200$ keV \citep{mcclintock2001,chaty2003}. The luminosity at $1<E<160$ keV was $L_{\rm{c}} = 1.31 \times 10^{36}$ erg s$^{-1}$ \citep{mcclintock2001}. 

According to \citet{maitra2009}, the 2005 outburst was fainter, with a steeper power-law in the X-rays. For this event, we consider a coronal luminosity of $L_{\rm{c}} = 9.0 \times 10^{35}$ erg s$^{-1}$, and a power-law of index $\alpha \sim 1.78$. 

The values of $r_{\rm{in}}$, and $kT_{\rm{max}}$ determine the spectrum of the accretion disk, and they were varied to obtain the best-fit model in each case. The parameters $L_{\rm{c}}$, $\alpha$, and $E_{\rm{c}}$ are inferred from observations; the remaining parameters ($B_{\rm{c}}$, $n_{e,i}$) are determined by applying the model discussed in Sect. \ref{model}, and their values are presented in Table \ref{table1}, for both outbursts. 

In a corona characterized by these parameters, we inject populations of non-thermal particles, both electrons and protons. We adopt the standard index  of $\Gamma = 2$ for the particle injection function (see the previous section for a discussion on the incidence of the particle index on the polarization degree). There were no observations of the source during the outbursts at MeV energies. According to our model, this is the energy range where the synchrotron radiation of secondary pairs dominates the spectrum. Since the content of pairs is related to the hadronic content in the source, the parameters $\eta$, $q$, and $a$ cannot be completely determined without simultaneous observations at TeV energies. Nevertheless, they are constrained by the magnetic energy density available for accelerating particles by magnetic reconnection. We adopt $\eta=0.01$ and $q=0.1$; these are values within the allowed range of space parameters \citep[e.g.,][]{vieyro2012}. For the proton to lepton ratio we use $a=100$, as observed in galactic cosmic rays. Changes in these last three variables, however, do not affect the predictions made on the polarization degree. Observations in the MeV-TeV energy range could be used to put stronger constrains on these parameters during future outbursts.


\begin{table*}[tb]
    \caption[]{Main parameters of the corona of \xte during its outbursts.}
   	\label{table1}
   	\centering
\begin{tabular}{lcc}
\hline\hline
Parameters & Outburst  & Outburst \\ [0.01cm]
					 & 2000      & 2005\\ [0.01cm]
\hline 
$L_{\rm{c}}$: corona luminosity [erg s$^{-1}$]                          & $1.3\times 10^{36}$	 & $9.0\times 10^{35}$	\\
$kT_{\rm{max}}$:  maximum disk temperature [keV]								& $22.0$    & $45.0$	\\
$r_{\rm{in}}$:   inner disk radius [$r_{\rm{g}}$] 							& $160.0$ 	  & $50.0$               	\\
$r_{\rm{in}}/R_{\rm{c}}$: inner disk/corona ratio               & $0.9$				& $0.9$						\\
$\alpha$: 			X-ray spectrum power-law index    							& $1.75$		& $1.78$							\\
$E_{\rm{c}}$:   X-ray spectrum cut-off [keV]							& $200$     & $200$  					\\
$B_{\rm{c}}$: 	magnetic field [G]				 																& $2.6 \times 10^4$			& $1.1 \times 10^5$\\
$n_{i},n{e}$:   plasma density [cm$^{-3}$] 											& $1.3 \times 10^{11}$	& $2.4 \times 10^{12}$ \\
\hline  \\[-0.5cm]
\end{tabular}
\end{table*} 

\vspace{0.5 cm}


\subsection{Results}\label{results}

\begin{figure}[htb]
\centering
{\includegraphics[width=0.41\textwidth,keepaspectratio]{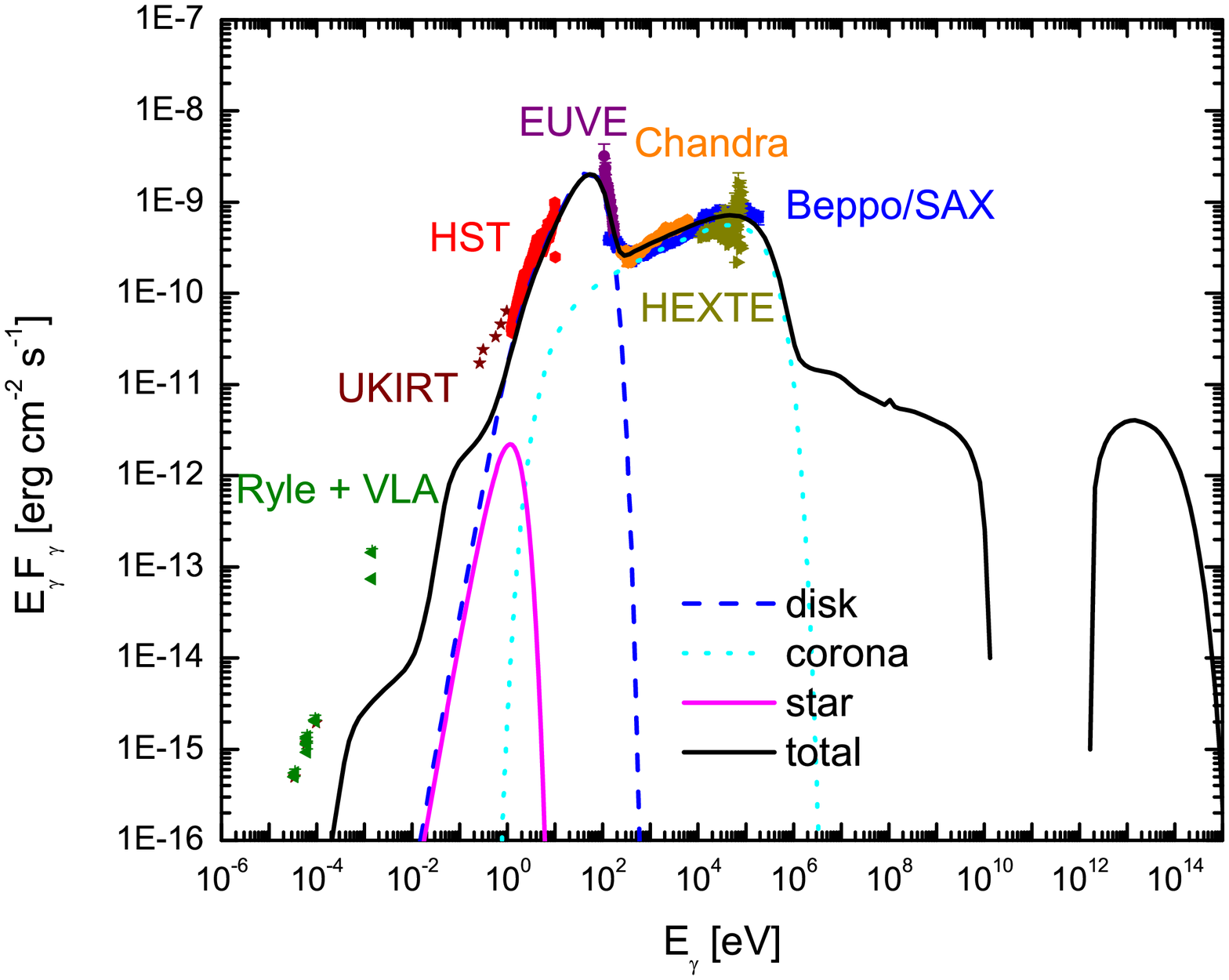}} \hspace{20pt} 
{\includegraphics[width=0.41\textwidth,keepaspectratio]{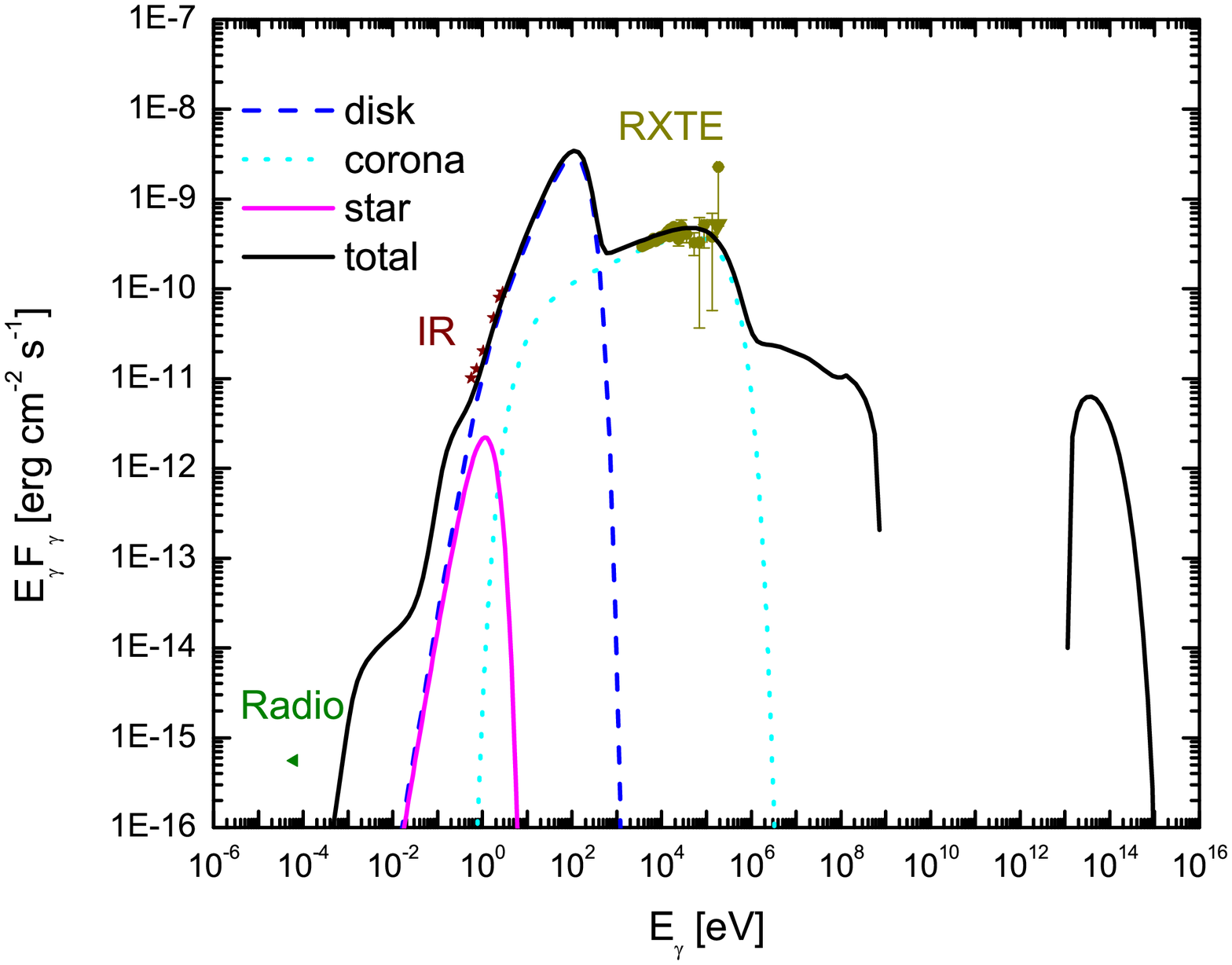}} \hfill 
\caption{Top panel: Computed SED for the 2000 outburst. Data from \citet{chaty2003}.Bottom panel: SED of the 2005 outburst. Data from \citet{brocksopp2010}.}
\label{fig:SED2}
\end{figure}

\begin{figure}[htb]
\centering
{\includegraphics[width=0.41\textwidth,keepaspectratio]{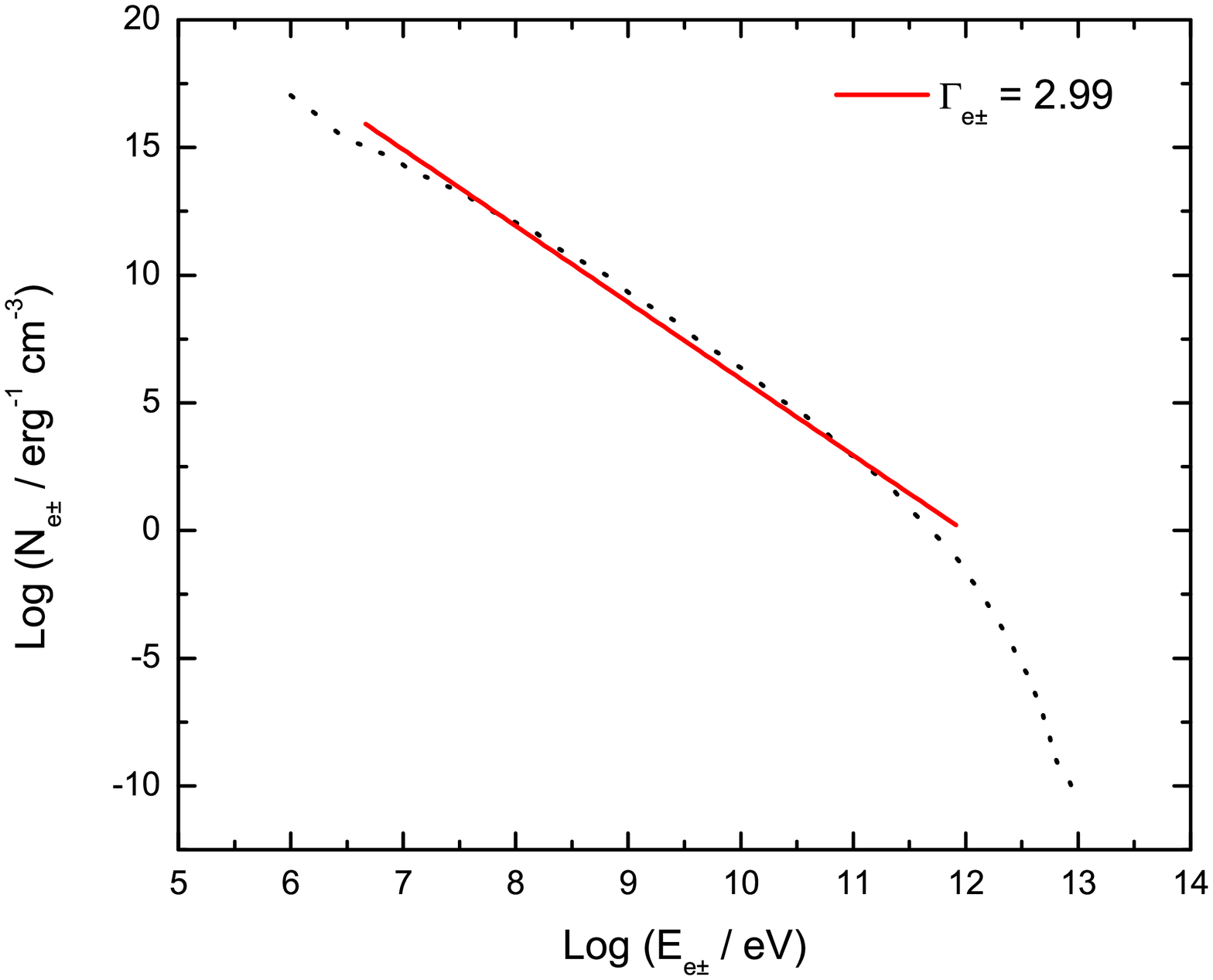}} \hspace{20pt} 
{\includegraphics[width=0.41\textwidth,keepaspectratio]{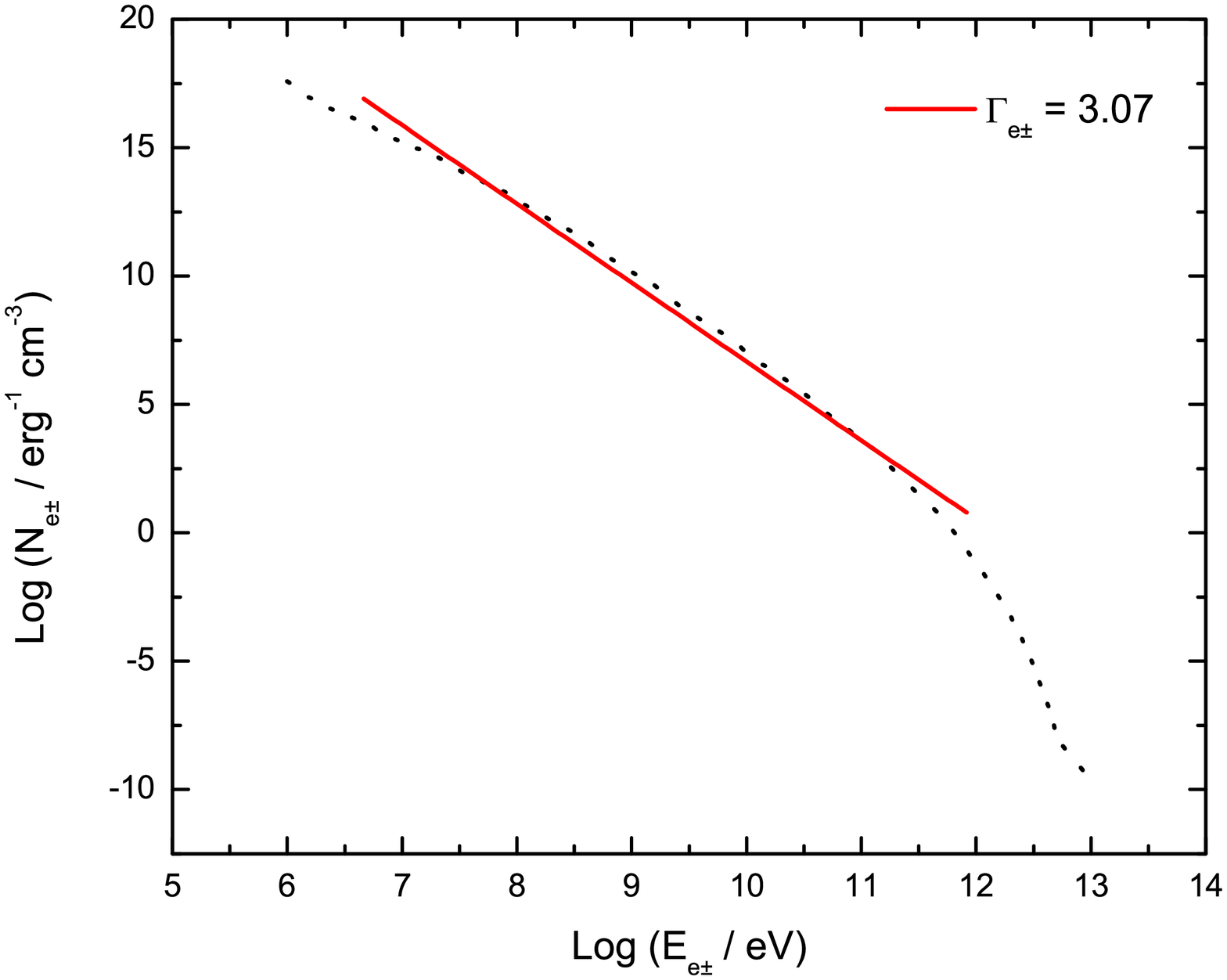}} \hfill 
\caption{Secondary pair distribution for the 2000 outburst (top panel) and the 2005 outburst (bottom panel). The solid line indicates the best power-law fit.}
\label{fig:pair2}
\end{figure}

Figure \ref{fig:SED2} shows the SEDs for the two outburst of \xtee, together with the data from different instruments. The corresponding final secondary pair distribution are shown in Fig. \ref{fig:pair2}. In both cases, the index of the pair distribution results $\sim 3$. Notice that the radio emission in our model is produced in the jet, and hence not calculated here.

The magnetic field at the base of the jet, $B_{\rm{jet}}(z_{0})$, was estimated by \citet{vila2012} for both outbursts. The estimated values are: $B_{\rm{jet}}(z_{0})=1.3\times 10^7$ G, $7.9\times 10^6$ G, with $z_{0}=50r_{\rm{g}}$, for the 2000 and 2005 outbursts, respectively. In order to obtain a value for the magnetic field in the disk inner radius, we extrapolate the results of \citet{vila2012} using the following expression:


\begin{equation}
B_{\rm{d}} = B_{\rm{jet}}(z_{0}) \Big( \frac{z_{0}}{r_{\rm{in}}} \Big)^{5/4},
\end{equation}

\noindent  where the dependence $B_{\theta} \propto (R_{\rm in}/r)^{5/4}$ is valid for the only non-null component of the magnetic field on the plane of the disk (i.e., $\theta = \pi/2$).


Finally, we compute the degree of polarization with the model presented in Sect. \ref{pol}, and obtain $\sim 23$\% for both outbursts.

The degree of polarization obtained from Eq. \ref{eq:ptot} depends mainly on the intrinsic properties of the synchrotron radiation. The geometry of the magnetic field only affects it through the factor $\left\langle \Delta B^2\right\rangle= \left\langle B_{\perp}^2 \right\rangle-\left\langle B_{\parallel}^2 \right\rangle$, as it was discussed in Sect. \ref{pol}.

A quantity that is strongly affected by the configuration of the magnetic field is the angle of polarization vector $\chi_{E}$. To obtain the Stokes parameters $Q$ and $U$,  we follow the method presented in \citet{nalewajko2012}. 

We adopt a Cartesian coordinate system $(x, y, z)$, in which the accretion disk axis (and consequently the magnetic field symmetry axis) is oriented along the $z$-axis. In this system, the unity vector $\vec{\breve{k}}=(\sin i, 0 , \cos i)$ determines the direction to the observer. An orthogonal system can be defined in the image plane by the unity vectors:

\begin{equation}
\begin{aligned}
\vec{\breve{v}}&=(\cos i, 0, -\sin i),\\
\vec{\breve{w}}&=(0,1,0).
\end{aligned}
\end{equation}

\noindent The Stokes parameters are then obtained as:

\begin{equation}
\begin{aligned}
Q  &= I P_{\rm tot} \cos(2 \chi_{E}) = \frac{C P_{\rm tot}}{4 \pi} \int d\Omega   \left[( \vec{B} \cdot \vec{\breve{w}} )^2 - (\vec{B} \cdot \vec{\breve{v}})^2 \right], \\
U  &= I P_{\rm tot} \sin(2 \chi_{E}) = \frac{C P_{\rm tot}}{4 \pi} \int d\Omega [-2(\vec{B} \cdot \vec{\breve{w}}) (\vec{B} \cdot \vec{\breve{v}}) ], 
\end{aligned}
\end{equation}

\noindent where $C$ is a constant that depends on the intensity $I$ of synchrotron radiation. Using the Cartesian components of the magnetic field resulting from Eq. \ref{eq:magneticField}, we obtain $U=0$ and $Q<0$; this implies $\chi_{E}=90^{\circ}$, as expected since the angle is measured from the projected direction of the magnetic field symmetry axis.

%


\section{Discussion}\label{discussion}

The degree of polarization that we obtained here for \xte ($\sim 21$ \%) is significantly lower than the one measured in Cygnus X-1 ($67-76$ \%; \citealt{laurent2011,jourdain2011}). The fact that the accretion rate onto the black hole of Cygnus X-1 is very stable over years (the observations by INTEGRAL covered a period of $6.5$ years), might be the reason of the high degree of polarization in this source \citep{russell2014}. Then, we should expect that a transient source as \xtee, where the magnetic field cannot reach such an ordered configuration, presents a lower polarization degree than Cygnus X-1.

The mechanism responsible of the presence of non-thermal particles in coronae is thought to be magnetic reconnection. Fast magnetic reconnection events have been studied for several structures of the disks, resulting in the acceleration of particles up to relativistic energies \citep[e.g.,][]{bette2010,singh2015,khiali2015}. The basic idea is that a first-order Fermi mechanism takes place within the reconnection zone caused by two converging magnetic fluxes of opposite polarity \citep{bette2005}. The resulting injection function of relativistic particles is a power-law with an index in the range of $1 \leq \Gamma \leq 3$ \citep{drury2012}. The value of the spectral index is related to the compression of the plasma, where a higher compression leads to a harder injection function. Given the dependence of the polarization degree of synchrotron radiation on the particle index (Eq. \ref{eq:pol}), a softer particle distribution should result in a higher degree of polarization. 

Changes in the accretion rate are related to the spectral state of the source. The low-hard state takes place for low values of $\dot{m}$; for higher values, the source switches to a high state, in which our model predicts a higher degree of polarization. This is mainly because of two factors: on the one hand, the disk extends up to the last stable orbit in the high-soft state, so the ordered magnetic field covers a higher volume fraction of the corona (i.e., $\xi$ increases). On the other hand, we obtain that for a high inclination of the lines of the magnetic field of the disk, the degree of polarization decreases. Then, the polarization degree should be lower in the low-hard than in the high-soft state. This is the opposite result of what should be expected if the polarized gamma radiation had its origin in a relativistic jet, since jets are not produced in the high-soft state. In addition, the angle of polarisation is also expected to change in the high-soft state. The magnetic field lines are inclined in the coronal region, so the polarisation should also change its direction from perpendicular (as obtained in Sect. \ref{results}, $\chi_{E}=90^{\circ}$) to parallel ($\chi_{E}=0^{\circ}$).  

There is no dedicated gamma-ray polarimeter currently in space, but different instruments that could test the predictions of our model are planned for a near future. This is the case of the Soft Gamma-ray Detector (SGD) onboard \textit{ASTRO-H}, an instrument that will be sensitive to $<10$\% polarization in the $50-200$ keV energy band, and it is planned to be launched in 2015 \citep{tajima2010}. Also, the missions of ESA, DUAL and GRI (Gamma-Ray Imager), include high energy polarimetric observations, and the CNRS (French National Research Agency) proposes to build a gamma-ray detector and polarimeter in the MeV - GeV energy range (project \textit{HARPO}, standing for Hermetic Argon Polarimeter). In addition, the \textit{ASTROGAM} space project is designed to study both steady and transient sources in the $0.3$ MeV - $1$ GeV energy range; it will represent an improvement of the sensitivity of a factor 10-30 in the range 0.3-30 MeV with respect to previous missions. It is also planned to surpass Fermi's sensitivity.  This mission will be equipped with polarimetric facilities at $E>$ MeV \citep{tavani2015}.

\section{Conclusions}\label{conclusions}

We have studied the polarization of synchrotron radiation of electron/positron pairs in the corona of a BHB in the low-hard state. In this state, the magnetic field of the disk is able to launch an outflow. We adopted a specific angular distribution for this ordered field, and we also estimated the random magnetic field of the corona. We applied our model to the source \xte, whose unique set of data allows us to accurately test our models and constrain polarization predictions. Taking into account the two contributions to the total magnetic field, we obtained a degree of polarization of $\sim 21$ \% for a state with similar characteristics to those observed in the 2000 and 2005 outbursts of \xtee. According to our model, if the source were observed again in outburst, it should present polarization in the energy range of $0.1-10$ MeV, detectable by future instruments.

At the time \xte was observed, there was no dedicated instruments operating at MeV energies. Our model predicts polarized MeV emission and simultaneous unpolarized TeV radiation for this source. Since this galactic source is located in a privilege low-absorption region and undergoes relatively frequent outbursts, the present quantitative predictions might be a valuable tool to put a coronal non-thermal model to the test through future multi-wavelength observations.

The presence of polarized gamma-ray radiation in the high state of XRBs would strongly support our model over those proposing the origin of the MeV tail emission in the jet.


\section*{Acknowledgments}

F.L.V. acknowledges Carolina Pepe and Ileana Andruchow for helpful discussions. This work was supported by the Argentine Agency ANPCyT (PICT 00878), as well as by grant AYA2013-47447-C3-1-P (Spain). S.C. and F.L.V. acknowledge funding by the Sorbonne Paris Cit\'e (SPC), Scientific Research Project - Argentina 2014. SC acknowledges funding from the Centre National d'Etudes Spatiales (CNES). This work is based on observations obtained with MINE (the Multi-wavelength {\it INTEGRAL} NEtwork), supported by the CNES.

\bibliographystyle{aa}
\bibliography{myrefs6}   

\end{document}